\documentclass[journal=jctcce,manuscript=article,layout=twocolumn]{achemso}

\usepackage[T1]{fontenc} % Use modern font encodings
\usepackage{graphicx}% Include figure files
\usepackage{dcolumn}% Align table columns on decimal point
\usepackage{bm}% bold math
\usepackage{subcaption}
\usepackage{siunitx}
\usepackage[version=4]{mhchem}
\usepackage{mathtools}
\usepackage{cuted}
\usepackage{hyperref}

\setkeys{acs}{doi = true, email = false}

\DeclareSIUnit\angstrom{\text{\AA}}

\author{Lars Dammann\footnote{\href{mailto:lars.dammann@tuhh.de}{lars.dammann@tuhh.de}}\,}
\affiliation[SoftMatterModeling]{Institute for Soft Matter Modeling, Hamburg University of Technology, 21073 Hamburg, Germany}
\alsoaffiliation[MaterialsX-rayPhysics]{Institute for Materials and X-Ray Physics, Hamburg University of Technology, 21073 Hamburg, Germany}
\alsoaffiliation[DESY]{Deutsches Elektronen-Synchrotron DESY, 22607 Hamburg, Germany}
\alsoaffiliation[SurfaceScience]{Institute of Surface Science, Helmholtz-Zentrum Hereon, 21502 Geesthacht, Germany}

\author{Richard Kohns}
\affiliation[MaterialsX-rayPhysics]{Institute for Materials and X-Ray Physics, Hamburg University of Technology, 21073 Hamburg, Germany}
\alsoaffiliation[DESY]{Deutsches Elektronen-Synchrotron DESY, 22607 Hamburg, Germany}

\author{Patrick Huber}
\affiliation[MaterialsX-rayPhysics]{Institute for Materials and X-Ray Physics, Hamburg University of Technology, 21073 Hamburg, Germany}
\alsoaffiliation[DESY]{Deutsches Elektronen-Synchrotron DESY, 22607 Hamburg, Germany}

\author{Robert H. Meißner\footnote{\href{mailto:robert.meissner@tuhh.de}{robert.meissner@tuhh.de}} }
\affiliation[SoftMatterModeling]{Institute for Soft Matter Modeling, Hamburg University of Technology, 21073 Hamburg, Germany}
\alsoaffiliation[SurfaceScience]{Institute of Surface Science, Helmholtz-Zentrum Hereon, 21502 Geesthacht, Germany}

\title{Maximum entropy mediated liquid-to-solid nucleation and transition}

\abbreviations{RDF,MD,MAE,WAXS,ADF,RMC,HRMC,LAMMPS,CNT}

\begin{document}

\begin{tocentry}
\includegraphics{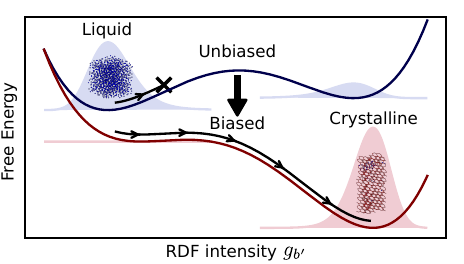}
\end{tocentry}

\begin{abstract}
  Molecular Dynamics (MD) simulations are a powerful tool for studying matter at the atomic scale. However, to simulate solids, an initial atomic structure is crucial for the successful execution of MD simulations, but can be difficult to prepare due to insufficient atomistic information. At the same time Wide Angle X-ray Scattering (WAXS) measurements can determine the Radial Distribution Function (RDF) of atomic structures. However, the interpretation of RDFs is often challenging. Here we present an algorithm that can bias MD simulations with RDFs by combining the information of the MD atomic interaction potential and the RDF under the principle of maximum relative entropy. We show that this algorithm can be used to adjust the RDF of one liquid model, e.g., the TIP3P water model, to reproduce the RDF and improve the Angular Distribution Function (ADF) of another model, such as the TIP4P/2005 water model. In addition, we demonstrate that the algorithm can initiate crystallization in liquid systems, leading to both stable and metastable crystalline states defined by the RDF, e.g., crystallization of water to ice and liquid \ce{TiO2} to rutile or anatase. Finally, we discuss how this method can be useful for improving interaction models, studying crystallization processes, interpreting measured RDFs, or training machine learned potentials.
\end{abstract}

\section{\label{sec:introduction}Introduction}
Understanding the structure of solid matter at the atomic level is an important task in materials science and a prerequisite for the development of new tailored materials. 
To this end, Molecular Dynamics (MD) simulations are an important method to study materials at the atomic scale. A crucial component to perform a MD simulation are suitable models of the interatomic potentials. However, knowledge of the interatomic potential alone is not sufficient to predict the structure of stable or metastable solid atomic structures. Therefore, the atomic structure of interest must be provided as another essential requirement to perform a physically meaningful MD simulation.

Sometimes it is not possible to provide the required atomic structure ab initio, because it is difficult to generate, e.g., in the case of amorphous structures, or not enough details are known about the exact atomic configuration. In this case an alternative option is to generate the target structure during the MD simulation from an initially unstable random or liquid phase. For this strategy to be successful, it is sometimes necessary to induce a phase transition in the initial system. This requires the system to move from one local free energy minimum to another stable free energy minimum by overcoming a free energy barrier. To date, several methods have been developed that can be used to overcome the free energy barrier and generate realistic solid atomic configurations during a MD simulation.

Some examples of these methods include crystal seed insertion \cite{piaggi_homogeneous_2022,pedevilla_heterogeneous_2018,sosso_ice_2016,sanz_homogeneous_2013,espinosa_seeding_2016}, simulated annealing \cite{biswas_simulated_1986,zhang_investigating_2023}, the introduction of a bias potential in one form or another \cite{piaggi_phase_2021,cheng_gibbs_2017,niu_molecular_2018,giberti_metadynamics_2015,bonati_silicon_2018,gobbo_nucleation_2018,zhang_improving_2019,nada_pathways_2020}, e.g., umbrella sampling \cite{torrie_nonphysical_1977} or well-tempered metadynamics \cite{barducci_well-tempered_2008}, 
or replica exchange methods \cite{coasne_freezing_2006,hung_molecular_2005,scalfi_gibbsthomson_2021,malolepsza_isobaric_2015,malolepsza_generalized_2015,malolepsza_water_2015}. 
However, each strategy has its own challenges. The introduction of a crystal seed requires a seed of the correct crystalline structure, which is problematic if the correct crystalline structure is not available. Introducing a meaningful bias potential requires the choice of a set of collective variables that summarize the position of the system in the free energy landscape. The collective variables usually have to be developed and chosen individually for each problem.
In contrast to these methods, simulated annealing and replica exchange methods are structure agnostic. However, the methods cannot be used to target a specific stable or metastable state of interest, since simulated annealing is designed to find the state of the global energy minimum and replica exchange methods are undirected with respect to the sampled configurations. In addition, it can be quite challenging to find the right parameters for the successful application of simulated annealing, biasing methods, or replica exchange methods. 

A common approach to the determination of unknown atomic structures is an experimental investigation with Wide Angle X-ray Scattering (WAXS). WAXS allows the determination of radial distribution functions (RDF) $g(r)$ of an atomic structure by Fourier transforming the measured isotropic total scattering structure function $S(q)$, which depends on the magnitude of the reciprocal scattering vector $q=\frac{2\pi}{r}$ \cite{egami_underneath_2012}. However, the interpretation of the atomic structure function $S(q)$ or the derived RDF $g(r)$ without any prior structural knowledge is an ill-posed inverse problem, since the generated RDF could originate from different atomic structures all resembling the same $S(q)$.

Considering the potential struggle to prepare atomic structures in MD simulations and to determine unknown atomic structures from experimental approaches, the question arises whether the information from the RDF can be combined with the information contained in the atomic interaction potentials used in MD simulations to generate atomic structures. 
For example, the Reverse Monte Carlo (RMC) method \cite{mcgreevy_reverse_1988,mcgreevy_reverse_2001,keen_reverse_2005} attempts to reconstruct the atomic structure from the atomic structure function $S(q)$ alone with unquantifiable fidelity. The Hybrid Reverse Monte Carlo (HRMC) method and its extension incorporating MD sampling \cite{bousige_optimized_2015,farmahini_hybrid_2015} on the other hand aims to improve the RMC algorithm by combining it with a molecular sampling scheme to reproduce the RDF $g(r)$ or the structure factor function $S(q)$. 
In HRMC, the RMC algorithm is added as a bias to the potential energy of the system. However, the mathematical structure of the introduced bias not only constrains the ensemble mean of the system to reproduce the RDF or the structure factor function, but it also constrains the variance of how the target function $g(r)$ or $S(q)$ can be reproduced from the system ensemble \cite{pitera_use_2012}. Since information about the variance of the objective function from an atomistic point of view is usually not available from experiments, this additional constraint is not supported by the experimental data. Furthermore, the application of the HRMC method requires the specification of a coupling parameter that determines the strength of the bias on original potential energy of the system. The choice of the coupling parameter can drastically alter the simulation results and should be based on the level of confidence in the original atomic potential or the influence of the experimentally determined data. Proper estimation of this trade-off makes the choice of the appropriate coupling parameter a difficult task. Therefore, a method that can incorporate the information contained in an RDF into MD simulations in a way that avoids inappropriate biases of the original force fields and facilitates the choice of appropriate coupling parameters would be an important step forward.

A suitable approach to combine information from measured observables with a thermodynamic system is the application of the principle of maximum entropy, first introduced by \citet{pitera_use_2012} and subsequently reviewed for the related principle of maximum relative entropy in the context of atomistic modeling by \citet{cesari_using_2018}. Prior to our work, relative entropy applications have been used in the development of coarse-graining potentials \cite{shell_relative_2008,shell_coarse-graining_2016,chaimovich_coarse-graining_2011} and the improvement of existing force fields \cite{cesari_combining_2016, frohlking_toward_2020}, while other works have shown that it is possible to reproduce target RDFs during molecular simulations using the principle of maximum relative entropy \cite{cilloco_relative_1993,white_efficient_2014,white_designing_2015}. In this work, we use the principle of maximum relative entropy to develop and implement an algorithm that influences the interaction potentials in MD simulations to reproduce target RDFs in liquid systems. Furthermore, we show that it can be sufficient to bias liquid systems with RDFs from stable or metastable crystalline states to induce a liquid-solid phase transition to a corresponding state that is defined by the target RDF.

\section{\label{sec:theory}Theory}
\subsection{\label{sec:rdf}Radial Distribution Functions}
\begin{figure}[!htb]
\includegraphics{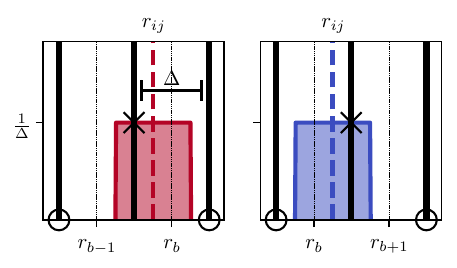}
\caption{\label{fig:histsketch} Sketch of the contribution of each distance $r_{ij}$ to the RDF $g_b$. The colored dashed lines represent the original delta distribution $\delta(\rho_b - r_{ij})$ at $r_{ij}$. We replace the delta distributions with rectangular kernel functions $\Theta(\rho_b - r_{ij})$ centered around $r_{ij}$. The contribution of the distance $r_{ij}$ to the bin $b$ is defined by the area under the rectangular function in the bin $b$ (lighter color). 
Due to the extent of the rectangular kernel function, the distance $r_{ij}$ contributes to the RDF intensity of two adjacent bins. If it is situated to the left of the center of the bin $b$, the affected bins are $b-1$ and $b$ (red). If it is situated to the right of the center of bin $b$, the affected bins are $b$ and $b+1$ (blue). 
The black cross indicates the non-zero value of $\Theta(r_b - \frac{\Delta}{2} - r_{ij}) = \Theta(r_{b-1} + \frac{\Delta}{2} - r_{ij})=\frac{1}{\Delta}$ (red) and $\Theta(r_b + \frac{\Delta}{2} - r_{ij}) = \Theta(r_{b+1} - \frac{\Delta}{2} - r_{ij})=\frac{1}{\Delta}$ (blue) at the bin boundaries, while the kernel function is zero at the other boundaries of the bins (black circle).}
\end{figure}
We can describe the radial distribution function of a macroscopic physical system with a large number of identical atoms $\tilde{N} \sim 10^{23}$ as a histogram $\tilde{g}_b$ with bins $b$ ranging from $1, \ldots, B$. Each bin extends over a range of $\pm\Delta/2$ around the radial position $r_b$. $\rho_b$ describes the position in bin $b$ and is defined on the interval $[r_b-\frac{\Delta}{2}, r_b+\frac{\Delta}{2}]$. Considering the atom coordinates $\mathcal{R}=\{\mathbf{r}^{\tilde{N}}\}$ the intensity of the radial distribution function $\tilde{g}_b$ in bin $b$ can be calculated by counting the weighted number of distances $r_{ij} = \lVert \mathbf{r}_{ij} \rVert = \lVert \mathbf{r}_i - \mathbf{r}_j \rVert$ between the individual particles $i$ and $j$. Mathematically, this can be represented as an integration over all delta distributions $\delta(\rho_b - r_{ij})$ positioned at each distance $r_{ij}$ situated within the bin boundaries from $\left(r_b-\frac{\Delta}{2}\right)$ to $\left(r_b+\frac{\Delta}{2}\right)$. Hence, $\tilde{g}_b$ reads as
\begin{equation}
    \label{eq:defRDF}
    \begin{split}
        \tilde{g}_b = &\frac{\tilde{V}}{\tilde{N}(\tilde{N}-1) \delta V_b}\\ 
        &\times \sum^{\tilde{N}}_{i=1} \sum^{\tilde{N}-1}_{j \neq i} \int_{\left(r_b-\frac{\Delta}{2}\right)}^{\left(r_b+\frac{\Delta}{2}\right)} {\rm d} \rho_b \, \delta(\rho_b - r_{ij}).            
    \end{split}
\end{equation}
Here, $\tilde{V}$ is the system volume and
\begin{equation}
    \delta V_b = \frac{4}{3} \pi \left[\left(r_b+\frac{\Delta}{2}\right)^3 - \left(r_b-\frac{\Delta}{2}\right)^3\right]
\end{equation}
the volume of a spherical bin. If the RDF is determined in a MD simulation, the number of atoms in the simulation $N$ and the volume $V$ of the simulated system will be significantly smaller, i.e. $N \ll \tilde{N}$ and $V \ll \tilde{V}$, than in a real macroscopic system. To simulate the RDF of a macroscopic system, the instantaneous RDFs as determined from the MD simulation $g_b$ are averaged $\langle\cdot\rangle$ over several time steps to approximate the macroscopic ensemble in the ergodic limit
\begin{equation}
    \label{eq:averageRDF}
    \begin{split}
        \tilde{g}_b &\approx \left\langle g_b
        \right\rangle\\
        &= \left\langle \frac{V}{N(N-1) \delta V_b} \right.\\
        &\quad \left. \times \sum^{N}_{i=1} \sum^{N-1}_{j \neq i} \int_{\left(r_b-\frac{\Delta}{2}\right)}^{\left(r_b+\frac{\Delta}{2}\right)} {\rm d} \rho_b \, \delta(\rho_b - r_{ij}) \right\rangle.        
    \end{split}
\end{equation}

To use the RDF as a bias potential in MD simulations, it has to be differentiable with respect to the interatomic distance $r_{ij}$. The non-differentiable $\delta$-distributions are thus replaced with differentiable kernel functions $\Theta(\rho_b - r_{ij})$ as an approximation to the original RDF.
The kernel function is centered around $r_{ij}$ and extends over the entire radial bin range $\pm\Delta/2$. Furthermore, the integral over the kernel function has to be equal to one.
In order to avoid any unjustified variations in the density of the contributions from $r_{ij}$ to the RDF over the bin range $\Delta$, we choose a simple rectangular function as kernel function. Thus, $\Theta(\rho_b - r_{ij})$ is defined as
\begin{equation}
  \Theta(\rho_b - r_{ij}) = 
    \begin{cases}
      \frac{1}{\Delta} & \text{if $\left|\frac{\rho_b-r_{ij}}{\Delta}\right| < \frac{1}{2}$}\\
      0 & \text{otherwise}
    \end{cases}.
\end{equation}
Due to the extent $\Delta$ of the kernel function, the contribution of $r_{ij}$ to the histogram is spread over two adjacent bins, if $r_{ij}$ is not positioned in the exact center $r_b$ of the bin $b$. The position of the adjacent bin to which the kernel function contributes to is dependent on the position of $r_{ij}$ in the bin.
If $r_b-\frac{\Delta}{2} < r_{ij} < r_b$ the kernel function contributes to the bins $b-1$ and $b$, while if $r_b < r_{ij} < r_b+\frac{\Delta}{2}$ the kernel function contributes to the bins $b$ and $b+1$ (see Figure~\ref{fig:histsketch}). After replacing the delta distributions with the rectangular functions eq~\eqref{eq:averageRDF} reads as
\begin{equation}
    \label{eq:approxRDF}
    \begin{split}
        \tilde{g}_b &\approx \left\langle g_b \right\rangle\\
        &= \left\langle \frac{V}{N(N-1) \delta V_b} \right.\\ 
        &\quad \left. \times \sum^{N}_{i=1} \sum^{N-1}_{j \neq i} \int_{\left(r_b-\frac{\Delta}{2}\right)}^{\left(r_b+\frac{\Delta}{2}\right)} {\rm d} \rho_b \, \Theta(\rho_b-r_{ij}) \right\rangle,
    \end{split}
\end{equation}
where the approximation holds for sufficiently large sample sizes of the $r_{ij}$ and small bin sizes $\Delta$.

Often a system contains several atom types. In this case, the total RDF can be described as the sum of the individual partial RDFs. Considering the atom types $\alpha$ and $\beta$, the partial radial distribution function, which describes the radial distribution function restricted to the atom pairs of type $\alpha$ and $\beta$, is written as $g_b^{\alpha \beta}$. To compute the total radial distribution function, the partial RDFs $g_b^{\alpha \beta}$ have to be weighted by the relative frequency $w^{\alpha \beta}$ of the $\alpha$ and $\beta$ pairs, resulting in
\begin{equation}
    \label{eq:partialRDF}
    g_b = \sum_{\alpha \beta} w_{\alpha \beta} g_b^{\alpha \beta}.
\end{equation}
If the total radial distribution function is measured by WAXS, the atom type specific structure factor $f$ has to be added to calculate the total RDF from its partial constituents. To mitigate the dependence of the structure factor from the length of the reciprocal scattering vector $q$, the Warren–Krutter–Morningstar approximation \cite{warren_fourier_1936} is used throughout this work. The total RDF is then given by
\begin{equation}
    \label{eq:scatteringRDF}
    g_b = \sum_{\alpha \beta} f_\alpha f_\beta w_{\alpha \beta} g_b^{\alpha \beta}.
\end{equation}

\subsection{\label{sec:maxrelent}Maximum relative entropy bias}
Equilibrium MD simulations are designed to sample particle coordinates from the generalized Boltzmann distribution \cite{gao_mathematics_2022}
\begin{equation}
        P_0(\mathcal{R}) = e^{-\beta \left(U(\mathcal{R}) - \sum^M_m  X_m x_m\right)}
\end{equation}
with the potential energy $U(\mathcal{R})$, the thermodynamic beta $\beta = \frac{1}{k_{\rm B} T}$, $M$ generalized forces $X_m$ (e.g., the pressure), and $M$ conjugate generalized coordinates $x_m$ (e.g., the volume).
The goal of this work is to bias the original generalized Boltzmann distribution $P_0$ of a MD simulation to obtain an updated ensemble distribution $P(\mathcal{R})$ that reproduces a target RDF $\left\langle g_b(\mathcal{R}) \right\rangle_{P} = \tilde{g}^{\rm target}_b$ in the MD simulation, while changing the original ensemble distribution $P_0$ ``as little as possible''.
We can translate this requirement into the goal of conveying as little information as possible from the target RDF $\tilde{g}^{\rm target}_b$ to the original interactions in the MD simulation, while still reproducing the target RDF. A biased ensemble distribution $P_{\rm ME}$ that satisfies these conditions can be found by maximizing the relative entropy $S$ \cite{caticha_relative_2004,pitera_use_2012}
\begin{equation}
    \label{eq:maximum-entropy}
    \begin{split}
        P_{\rm ME} &= \underset{P}{\textrm{arg\,max}} \, S[P||P_0] \\
        &= \underset{P}{\textrm{arg\,max}} \left[
        - \int \, {\rm d}\mathcal{R} P(\mathcal{R}) \ln{\frac{P(\mathcal{R})}{P_0(\mathcal{R})}} \right]    
    \end{split}    
\end{equation}
under the constraints
\begin{align}
    \label{eq:average}
    \int \, {\rm d}\mathcal{R} g_b(\mathcal{R}) P(\mathcal{R})&=\left\langle g_b(\mathcal{R}) \right\rangle_{P} = \tilde{g}_b^{\rm target}\\
    \label{eq:norm}
    \int \, {\rm d}\mathcal{R} P(\mathcal{R}) &= 1.
\end{align}
Here, $\int \, {\rm d}\mathcal{R} \cdot$ implies an integral over the whole phase space. This approach is called the principle of maximum relative entropy. A review on biasing with the maximum relative entropy approach is provided by~\citet{cesari_using_2018}. Solving the constrained optimization problem with the method of Lagrange multipliers leads to a posterior probability distribution that reads as
\begin{equation}
    \begin{split}
        P_{\rm ME}(\mathcal{R}) &\propto e^{- \sum^B_{b=1} \lambda_b g_b(\mathcal{R})} P_0(\mathcal{R})\\
        &= e^{- \boldsymbol{\lambda} \cdot \mathbf{g}(\mathcal{R})} P_0(\mathcal{R}).        
    \end{split}
\end{equation}
Thus, the generalized Boltzmann distribution is updated to be
\begin{equation}
    \label{eq:genforces}
    \begin{split}
        &P_{\rm ME}(\mathcal{R})\\
        &= e^{-\beta \left((U(\mathcal{R}) - \sum_m X_m x_m\right) + \sum_{b=1}^B \lambda_b g_b(\mathcal{R})}\\
        &= e^{-\beta \left((U(\mathcal{R}) - \sum_m X_m x_m + \sum_{b=1}^B \hat{\lambda}_b g_b(\mathcal{R})\right)}\\
        &= e^{-\beta \left(U(\mathcal{R}) - \sum_m X_m x_m + V(\mathcal{R})\right)}
    \end{split}
\end{equation}
with the linear bias potential
\begin{equation}
    \label{eq:biaspot}
    V(\mathcal{R}) = \sum_{b=1}^B \hat{\lambda}_b g_b(\mathcal{R}) = \boldsymbol{\hat{\lambda}} \mathbf{g}(\mathcal{R})
\end{equation}
and the definition that
\begin{equation}
    \hat{\lambda}_b = \frac{1}{\beta} \lambda_b.
\end{equation}
The similarity of the term $\sum_m X_m x_m$ to $\sum_{b=1}^B \hat{\lambda}_b g_b$ in eq~\eqref{eq:genforces} is due to the fact that thermodynamic systems reproduce an average of the generalized coordinates $\langle x_m \rangle$ specified by $X_m$ while maximizing the entropy of the distribution of generalized coordinates of the system. In this sense, the values $g_b(\mathcal{R})$ can be compared to ``generalized coordinates'' and $\hat{\lambda}_b$ to ``generalized forces''.

In MD simulations, the bias potential $V(\mathcal{R})$ adds a force $\mathbf{F}_i$ to the atom $i$, which can be expressed as
\begin{equation}
    \label{eq:biasforce}
    \mathbf{F}_i = \mathbf{\nabla}_{i} V(\mathcal{R}) = \sum_{b=1}^B \hat{\lambda}_b \mathbf{\nabla}_{i} g_b(\mathcal{R}).
\end{equation}
The derivative of the RDF contribution $\nabla_{i} g_b(\mathcal{R})$ with respect to the coordinate $\mathbf{r}_i$ of atom $i$ is
\begin{equation}
    \begin{split}
        &\nabla_{i} g_b(\mathcal{R})\\
        &= \frac{V}{N(N-1) \delta V_b}\\
        &\quad \times \sum^{N-1}_{j \neq i} \mathbf{\nabla}_{r_i} \int_{\left(r_b-\frac{\Delta}{2}\right)}^{\left(r_b+\frac{\Delta}{2}\right)} {\rm d} \rho_b \, \Theta(\rho_b-r_{ij})\label{eq:gradient}\\
        &= \frac{V}{N(N-1) \delta V_b} \frac{\mathbf{r}_{ij}}{r_{ij}}\\
        &\quad \times \sum^{N-1}_{j \neq i} \left[\Theta\left(r_b - \frac{\Delta}{2} - {r}_{ij}\right)\right.\\
        &\hspace{3.5em} \left. -\Theta\left(r_b + \frac{\Delta}{2} - {r}_{ij}\right)\right].
    \end{split}
\end{equation}
By inserting eq~\eqref{eq:gradient} into eq~\eqref{eq:biasforce} and rearranging, we get
\begin{equation}
    \label{eq:atomforce}
    \mathbf{F}_i = \sum^{N-1}_{j \neq i} \mathbf{F}_{ij} = \sum^{N-1}_{j \neq i} \frac{V \hat{F}_{ij}}{N(N-1)} \frac{\mathbf{r}_{ij}}{r_{ij}}
\end{equation}
with
\begin{equation}
    \label{eq:interactionforce}
    \mathbf{F}_{ij} = \frac{V \hat{F}_{ij}}{N(N-1)} \frac{\mathbf{r}_{ij}}{r_{ij}}
\end{equation}
and
\begin{equation}
    \label{eq:binforceall}    
    \begin{split}
        \hat{F}_{ij} &= \sum_{b=1}^B \hat{F}^{b}_{ij}\\
        &= \sum_{b=1}^B \frac{\hat{\lambda}_b}{\delta V_b} \left[\Theta\left(r_b + \frac{\Delta}{2} - {r}_{ij}\right) \right.\\
        &\hspace{4.25em} \left. -\Theta\left(r_b-\frac{\Delta}{2} - {r}_{ij}\right)\right].
    \end{split}
\end{equation}
Note that $\Theta\left(r_b - \frac{\Delta}{2} - {r}_{ij}\right)$ is the value of the kernel function centered at $r_{ij}$ at the lower boundary of bin $b$ and $\Theta\left(r_b + \frac{\Delta}{2} - {r}_{ij}\right)$ the value of the same kernel function at the upper boundary of bin $b$.
We can simplify eq~\eqref{eq:binforceall} by considering that each kernel function $\Theta$ around $r_{ij}$ contributes to only two adjacent bins, the bin $\hat{b}$ in which $r_{ij}$ is located and a neighboring bin $\hat{b}'$. Thus, the kernel function crosses the bin boundaries only at one bin boundary position (see Figure~\ref{fig:histsketch} at the location of the black cross). Since eq~\eqref{eq:binforceall} depends only on the values of the kernel function $\Theta$ at the bin boundaries, only two terms, stemming from $b=\hat{b}$ and $b=\hat{b}'$ in the sum $\sum_{b=1}^B$ over all bins in eq~\eqref{eq:binforceall} are non-zero, or more precisely $\frac{1}{\Delta}$ (black cross in Figure~\ref{fig:histsketch}), and contribute to the term eq~\eqref{eq:binforceall}. At the position of all other bin boundaries, the kernel function is zero and does \textbf{not} contribute to the term eq~\eqref{eq:binforceall} (black circles in Figure~\ref{fig:histsketch}). Whether $r_{ij}$ is positioned to the left or right of the bin center $r_{\hat{b}}$ determines whether the adjacent bin is $\hat{b}'=\hat{b}-1$ (red kernel function in Figure~\ref{fig:histsketch}) or $\hat{b}'=\hat{b}+1$ (blue kernel function in Figure~\ref{fig:histsketch}). So eq~\eqref{eq:binforceall} can be simplified to
\begin{equation}
    \label{eq:binforce}
    \begin{split}
        &\hat{F}_{ij} =\\
        &\begin{cases}
          \text{$\frac{1}{\Delta}\left[\frac{\hat{\lambda}_{\hat{b}}}{\delta V_{\hat{b}}} - \frac{\hat{\lambda}_{\hat{b}-1}}{\delta V_{\hat{b}-1}}\right]$} & \text{if $r_{\hat{b}}-\frac{\Delta}{2} < r_{ij} < r_{\hat{b}}$}\\
          \text{$\frac{1}{\Delta}\left[\frac{\hat{\lambda}_{\hat{b}+1}}{\delta V_{\hat{b}+1}} - \frac{\hat{\lambda}_{\hat{b}}}{\delta V_{\hat{b}}}\right]$} & \text{if $r_{\hat{b}} < r_{ij} < r_{\hat{b}}+\frac{\Delta}{2}$}\\
        \end{cases}.
    \end{split}
\end{equation}

If partial RDF compositions (eq~\eqref{eq:partialRDF}) or RDFs measured by X-ray scattering (eq~\eqref{eq:scatteringRDF}) are considered, the weights of the partial RDFs have to be included in the forces and eq \eqref{eq:interactionforce} is altered to
\begin{equation}
    \mathbf{F}_{ij} = w_{\alpha \beta} \frac{V \hat{F}_{ij}}{N(N-1)} \frac{\mathbf{r}_{ij}}{r_{ij}}    
\end{equation}
or
\begin{equation}
    \label{eq:scatterforce}
    \mathbf{F}_{ij} = f_{\alpha} f_{\beta} w_{\alpha \beta} \frac{V \hat{F}_{ij}}{N(N-1)} \frac{\mathbf{r}_{ij}}{r_{ij}},    
\end{equation}
respectively. Since the bias force is generated by a pairwise potential, it will contribute to the original virial pressure of the system, $T_{0}$, by adding a term, $T_{\rm ME}$. Consequently, the new total virial pressure of the system, $T_{\rm total}$, is given by
\begin{equation}
    T_{\rm total} = T_{0} + T_{\rm ME}.
\end{equation}
The virial resulting from the bias can be calculated as \cite{de_miguel_nature_2006}
\begin{equation}    
\label{eq:virial}
T_{\rm ME} = \frac{ 1 }{ V d } \sum^{N}_{i} \sum^{N-1}_{j \neq i} \mathbf{F}_{ij}  \mathbf{r}_{ij},
\end{equation}
where $d$ is the dimensionality of the system, e.g., three for a three-dimensional system. By substituting eq~\eqref{eq:interactionforce} and eq~\eqref{eq:binforceall}, eq~\eqref{eq:virial} can be transformed to
\begin{equation}
    T_{\rm ME} = \sum_{b=1}^B T^{b}_{\rm ME}.
\end{equation}
with
\begin{equation}
    \label{eq:virialperbin}
    T^{b}_{\rm ME} = \frac{ 1 }{ d N(N-1)} \sum^{N}_{i} \sum^{N-1}_{j \neq i} \hat{F}^{b}_{ij} r_{ij},
\end{equation}
where $T^{b}_{\rm ME}$ represents a measure for the bias induced virial per bin $b$.

\subsection{Determination of Lagrange multiplier \texorpdfstring{$\lambda_b$}{lambda b}}
To bias the original probability distribution $P_0(\mathcal{R})$, the parameters $\lambda_b$ need to be determined. Following \citet{pitera_use_2012} the function
\begin{equation}
    \label{eq:gamma}
    \begin{split}
        \Gamma(\boldsymbol{\lambda})
        = &\ln{\left[\int d\mathcal{R} P_0(\mathcal{R}) e^{-\sum^B_{b=1} \lambda_b g_b(\mathcal{R})}\right]}\\
        &+ \sum^B_{b=1} \lambda_b \tilde{g}_b^{\rm target}        
    \end{split}
\end{equation}
is introduced, which combines eq~\eqref{eq:maximum-entropy} with eq~\eqref{eq:average} and eq~\eqref{eq:norm} by a Legendre transformation \cite{mead_maximum_1984}. The set of parameters $\boldsymbol{\lambda}$ that solve the combined equations \eqref{eq:maximum-entropy}--\eqref{eq:norm} can be found by minimizing eq~\eqref{eq:gamma}.
Since the intensities in the bins of the RDF are in general statistically correlated, the function $\Gamma(\boldsymbol{\lambda})$ is not convex and there is no guarantee to find an optimal set of parameters $\boldsymbol{\lambda}$. 

In this study, we use gradient descent with step size of $\gamma$ to iteratively update the Lagrange multipliers during the simulation to minimize $\Gamma(\boldsymbol{\lambda})$. 
As discussed at the end of Sec.~\ref{sec:maxrelent}, the bias force leads to a change in the virial of the system (eq~\eqref{eq:virial}). Depending on the target RDF, this can lead to a significant change in volume in systems simulated as isobaric ensembles. 
To counteract an undesired large change in volume, we introduce an optional scaling parameter $\kappa_b$ into the gradient descent algorithm. The scaling parameter $\kappa_b$ adjusts the update of the Lagrange parameters $\lambda_b$ based on the average contribution of the induced viral per bin $\langle T^{b}_{\rm ME} \rangle$. To counteract excessive contraction or expansion of the system, we reduce the update rate by setting $\kappa_b < 1.0$ for bins $b$ for which the average virial per bin $\langle T^{b}_{\rm ME} \rangle$ contributes negatively or positively to the total virial, respectively. The update rates for the other bins remain unchanged with $\kappa_b = 1.0$. With the additional parameter $\kappa_b$ we observed satisfactory results in minimizing $\Gamma(\boldsymbol{\lambda})$ with gradient descent for all considered systems.

Each update of the parameters $\lambda_b$ is calculated as
\begin{equation}
    \label{eq:gradientdescent}
    \begin{split}
        \lambda_b^\prime &= \lambda_b + \gamma \kappa_b \frac{\frac{\partial \Gamma}{\partial \lambda_b}}{\sum_{b=1}^B \left|\frac{\partial \Gamma}{\partial \lambda_b}\right|}\\
        &= \lambda_b + \gamma \kappa_b \frac{\tilde{g}_b^{\rm target} - \langle g_b(\mathcal{R}) \rangle}{\sum_{b=1}^B \left|\tilde{g}_b^{\rm target} - \langle g_b(\mathcal{R}) \rangle\right|}.
    \end{split}
\end{equation}
The sum over all bins of the absolute values of the RDF in the denominator of the fraction ensures that the updates of the Lagrange multipliers $\lambda_b$ are constant, regardless of how similar the measured RDF $\langle g_b(\mathcal{R}) \rangle$ is to the target RDF $g^{\rm target}$. We have found that this facilitates the choice of a meaningful update parameter $\gamma$ that can be kept constant throughout the simulation.

\subsection{\label{sec:freeenergy}Potential of mean force}
\begin{figure}[htb]
\includegraphics{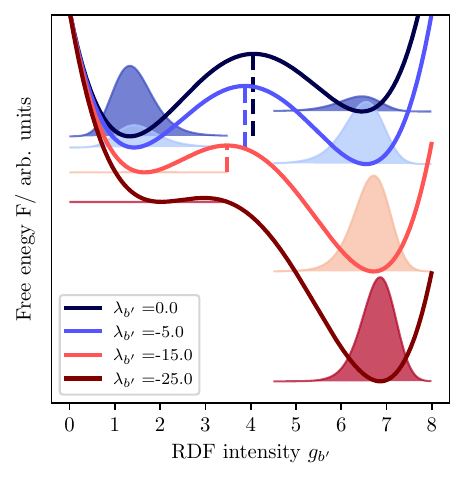}
\caption{\label{fig:pmf} Hypothetical free energy surface as a function of the RDF intensity $g_{b'}$ at one specific bin position $r_{b'}$. Two local free energy minima are visible that are separated by a free energy barrier. The resulting Boltzmann probability distribution is indicated by slightly lighter distributions in the energy minima. An increase of the bias is expressed by the decrease of the Lagrange multiplier $\lambda_{b'}$ from $\lambda_{b'}=0.0$ (dark blue) to $\lambda_{b'}=-25.0$ (dark red).
}
\end{figure}
The free energy or potential of mean force as a function of the instantaneous RDF obtained in the MD simulation $g_b$ forms a hypersurface
\begin{equation}
    \begin{split}
        F_0(\mathbf{g}) = &-\frac{1}{\beta} \log{\int \, d\mathcal{R} \delta(\mathbf{g}^{\prime}(\mathcal{R}) - \mathbf{g}) P_0(\mathcal{R})}
    \end{split}
\end{equation}
with as many dimensions as there are biased bins $B$. This hypersurface is tilted by the applied linear bias
\begin{equation}
    \begin{split}
        F(\mathbf{g}) &= -\frac{1}{\beta} \log{\int \, d\mathcal{R} \delta(\mathbf{g}^{\prime}(\mathcal{R}) - \mathbf{g}) P_{\rm ME}(\mathcal{R})}\\
        &= -\frac{1}{\beta} \log{\int \, d\mathcal{R} \delta(\mathbf{g}^{\prime}(\mathcal{R})} - \mathbf{g}) P_{0}(\mathcal{R}) e^{\boldsymbol{\lambda} \mathbf{g}^{\prime}(\mathcal{R})}\\
        &= F_0(\mathbf{g}) + k_B T \boldsymbol{\lambda} \mathbf{g}
    \end{split}
\end{equation}
as discussed in \citet{cesari_using_2018}.

Figure~\ref{fig:pmf} shows the hypothetical free energy surface for an exemplary one-dimensional case of a single bin $b'$. The unbiased system $\lambda_{b'} = 0.0$ (dark blue) has two local free energy minima. The state at $g_{b'} \approx 1.5$ is the lowest free energy state and thus stable, while the state at about $g_{b'} \approx 6.5$ is a metastable state of higher energy that is separated from the stable state by a free energy barrier indicated by the dashed line. The resulting Boltzmann probability distribution is indicated by slightly lighter distributions in the energy minima. The stepwise decrease of the Lagrange multiplier down to $\lambda_{b'} = -25.0$ (from dark blue to dark red) causes the free energy profile to tilt downward with increasing intensity of $g_{b'}$. The induced tilt has two important effects on the free energy surface. First, the free energy of the metastable state is reduced to the point where it becomes the new minimum energy state of the system (in Figure~\ref{fig:pmf} the new global minimum changes from $g_{b'} \approx 1.5$ to $g_{b'} \approx 7$). Thus, the former metastable state becomes the new stable state and vice versa. As a result, crystalline phases may become stable in the simulation at temperatures and/or pressures where the phase was not stable prior to the induced bias. Second, the free energy barrier is lowered. Reducing the barrier increases the probability of a state transition up to the point where, by increasing the bias, the free energy barrier disappears and a state transition is certain to occur. On multidimensional free energy surfaces with multiple bins $b$, the direction of tilt of the free energy surface determines the amount of energy reduction of the target state and the barrier. The tilt angle between the different dimensions defined by the bins $b$ in the presented algorithm is implicitly chosen by the step size $\gamma$ and  the virial weighting $\kappa_b$ at which each individual Langrange multiplier $\lambda_b$ is updated.

\section{\label{sec:results}Results and Discussion}
We implemented the maximum relative entropy formalism as an extension to the Large-scale Atomic/Molecular Massively Parallel Simulator (LAMMPS) \cite{thompson_lammps_2022}. To incorporate the atomic structure factor dependence of the RDFs, we build on the code of \citet{coleman_computational_2014}, which in turn uses an analytical approximation of the atomic structure factors \cite{prince_international_2007} parameterized by \citet{peng_robust_1996}. We then evaluated the implemented algorithm in two different applications. In the first application, we biased a liquid water system modeled with the TIP3P \cite{jorgensen_comparison_1983} model to reproduce the RDF of another system modeled with the TIP4P/2005 \cite{abascal_general_2005} water model. In the second application, we investigated the potential of the algorithm to moderate liquid-solid phase transitions. Therefore, we biased a liquid water system to reproduce the hexagonal phase I\textsubscript{h} and a liquid \ce{TiO2} system to reproduce the stable polymorphs rutile or anatase depending on of the applied target RDF. All examples presented were simulated under the assumption that the RDFs were generated from WAXS measurements. This means that atomic form factors had to be used to weight the partial RDF contributions in the RDF and the resulting forces (eq~\eqref{eq:scatteringRDF} and eq~\eqref{eq:scatterforce}). We chose WAXS target RDFs because they are the most complex to reproduce with the additional weighting factors. Therefore, we expect the formalism to work similarly for target RDFs generated by identical particles or partial RDF compositions. In the following sections, we will present the simulated RDFs $\langle g_b \rangle$ with small bin sizes $\Delta$. Thus, the bin positions $r_b$ and the RDFs $\langle g_b \rangle$ are represented as continuous $\{r_1,\ldots,r_B\} \approx r$ and $\{\langle g_1 \rangle,\ldots,\langle g_B \rangle\} \approx g(r)$ for clarity.

\subsection{\label{sec:water}Biasing of liquid systems: Water}
\begin{figure*}[!t]
    \centering
    \begin{subfigure}{.5\textwidth}
      \centering
      \includegraphics[width=1\linewidth]{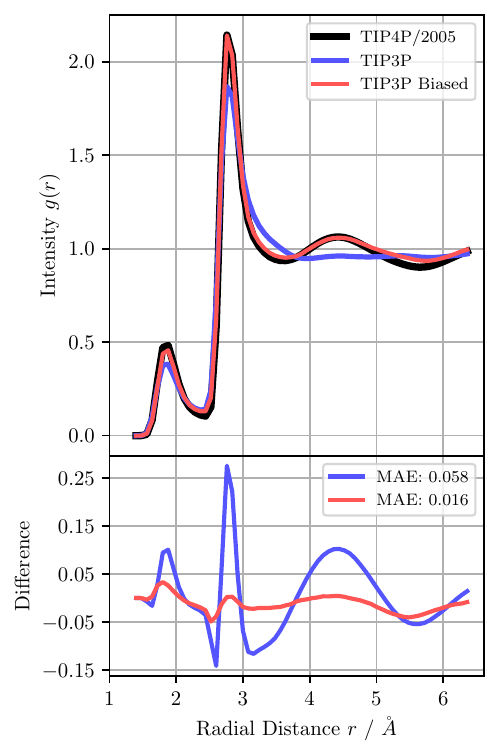}
      \caption{RDFs $g(r)$ and their differences.}
      \label{fig:liquid-rdf-difference}
    \end{subfigure}%
    \begin{subfigure}{.5\textwidth}
      \centering
      \includegraphics[width=1\linewidth]{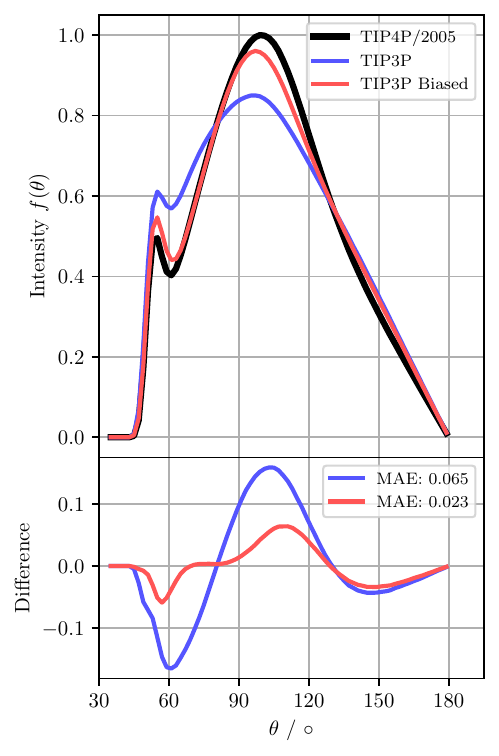}
      \caption{ADFs $f(\theta)$ and their differences.}
      \label{fig:liquid-adf-difference}
    \end{subfigure}%
    \caption{\label{fig:liquid-bias}Upper panel: RDF (left) and ADF (right) of liquid water systems modeled with different interatomic potentials. Black is the RDF/ADF of the TIP4P/2005 water model from which the target RDF was extracted. The RDF/ADF of the unbiased TIP3P water model is depicted in blue, while the biased TIP3P water model is depicted in red.
    Bottom panel: Difference in the RDF/ADF between the target TIP4P/2005 water model and the unbiased (blue) and biased (red) TIP3P water model. The two respective mean absolute errors (MAE) between the functions are given in the plot legend, with smaller values indicating better accuracy.}
\end{figure*}
A wide variety of atomic interaction potentials have been developed to reproduce different properties of water, each with a different level of computational effort \cite{kadaoluwa_pathirannahalage_systematic_2021}. Among other properties, MD simulations with different water models tend to have different radial distribution functions. Here, we demonstrate that our method can be used to fit the RDF of a liquid system modeled with one atomic interaction potential to the RDF of a liquid system modeled with a different atomic interaction potential. As a demonstration, we biased the atomic interaction potential of TIP3P \cite{jorgensen_comparison_1983} to reproduce the RDF of the TIP4P/2005 \cite{abascal_general_2005} model. 

The TIP3P water model is a three-sided model where the water molecule is simulated as three individual atoms. In contrast, the TIP4P/2005 water model is a four-sided model that adds an additional dummy atom to the water molecule. This dummy atom has no mass, but carries a charge, allowing for a more accurate representation of the charge distribution among water molecules. This makes the TIP4P/2005 model more suitable for reproducing a variety of bulk water properties, including density, viscosity, and diffusion coefficient, as well as the local water structuring \cite{dorrani_comparative_2023}. However, this advantage is offset by the increased computational complexity. 

The goal of this demonstration is to use the maximum entropy algorithm to improve the agreement between the TIP3P and TIP4P/2005 models by biasing the TIP3P model. For this purpose, we simulated a liquid water system with the TIP4P/2005 model at a temperature of \SI{300}{\kelvin} and a pressure of \SI{1}{atm}. From the equilibrated system, we calculated a target RDF $\tilde{g}^{\rm target}_{b}$ ranging from \SI{0}{\angstrom} to \SI{8}{\angstrom} with a bin size of \SI{0.08}{\angstrom}. This target RDF was then used to bias a TIP3P equilibrium system according to the maximum relative entropy formalism. The RDF $\langle g_b \rangle$ was calculated every ten time steps and averaged over 100 time steps to conduct the update of the Lagrange multiplier $\lambda_b$ with a step size of $\gamma=10.0$ and no virial specific weighting, i.e., $\kappa_b = 1$ for all bins.

Figure \ref{fig:liquid-rdf-difference} shows the resulting RDF from the TIP3P model before (blue) and after (red) the application of the bias compared to the target RDF from the TIP4P/2005 model (black) in the upper part of the figure. The lower part of the figure depicts the difference between the TIP4P/2005 target RDF and the unbiased and biased RDFs of the TIP3P model in blue and red, respectively. After applying the bias, the initial mean absolute error (MAE) between the RDFs was reduced from 0.058 to 0.016.

In addition to the effect of the bias on the generated RDFs, we also investigated the influence of the introduced bias on another characteristic of the model unrelated to the RDF, i.e., the angular distribution function (ADF) $f(\theta)$ of the oxygen atoms within a distance of \SI{0}{\angstrom} to \SI{3.4}{\angstrom} over an angle $\theta$ of $180^{\circ}$ in $2^{\circ}$ bins, as shown in Figure \ref{fig:liquid-adf-difference}. Similar to Figure \ref{fig:liquid-rdf-difference}, in the upper part of the figure the ADF of the target model TIP4P/2005 is shown in black, the unbiased TIP3P in blue and the biased TIP3P in red. In the lower part of the figure the difference between the ADF of the TIP4P/2005 model and the unbiased and biased TIP3P models is shown in blue and red, respectively. The ADF of the TIP3P model shows a clear change with the introduction of the bias. Similar to the RDF generated by the original TIP3P model, the unbiased ADF was less structured than the ADF generated by the TIP4P/2005 target model. With the application of the bias, the difference between the TIP4P/2005 and the biased TIP3P ADF decreased from a MAE of 0.065 to 0.023. 

Overall, the results suggest that the maximum relative entropy formalism can be a suitable approach to improve an atomic interaction model with a more sophisticated model or experimental measurement.

\subsection{Crystallization of water}
\begin{figure*}[!t]
\centering
    \includegraphics[width=1\linewidth]{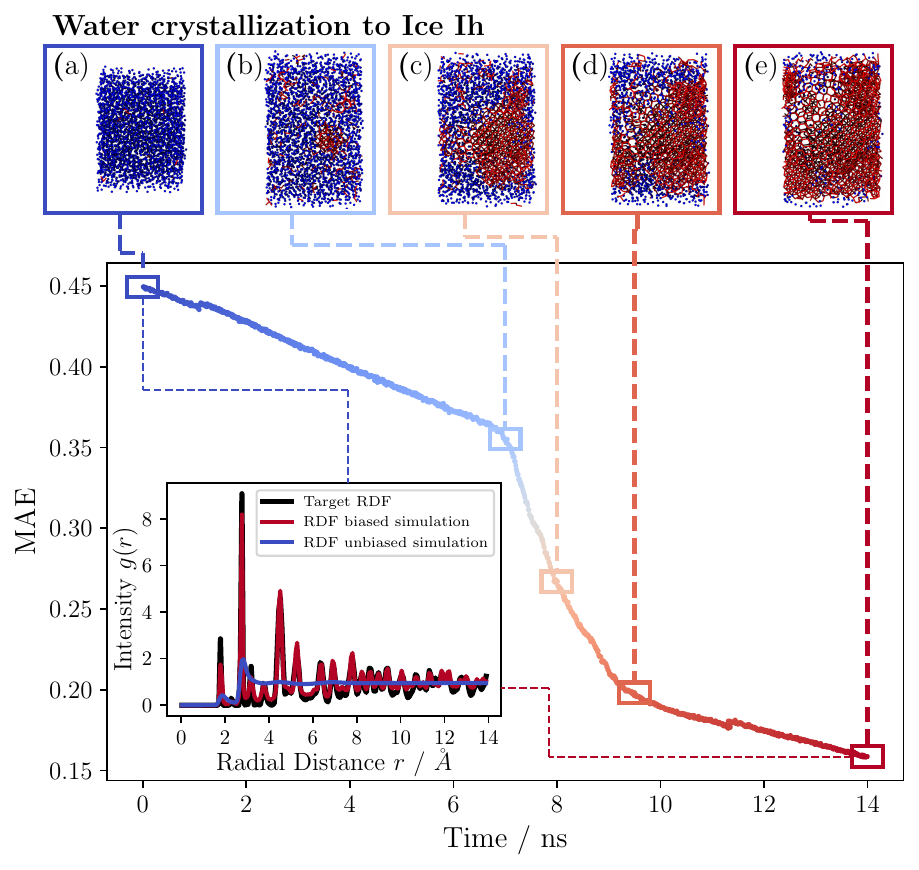}
    \caption{Bias-induced liquid-solid transition of a liquid water system modeled with TIP4P/ICE at \SI{360}{\kelvin} and \SI{1}{atm}. The mean absolute error (MAE) between the RDF of the biased water system and the target RDF of the simulated hexagonal ice I\textsubscript{h} is depicted. Several snapshots of the biased simulation show the state of the oxygen atoms in the system from left to right at \SI{0}{\nano\second}, \SI{7}{\nano\second}, \SI{8}{\nano\second}, \SI{9.5}{\nano\second}, and \SI{14}{\nano\second}. The atoms in the liquid and hexagonal ice phase are colored blue and red, respectively. The inlet shows the RDF of the unbiased liquid simulation (blue) and the final state of the system (red) compared to the target RDF (black).
    }
    \label{fig:ice} 
\end{figure*}
Homogeneous crystallization of water is a process of great interest, but is challenging to study with MD simulations. As predicted by classical nucleation theory (CNT), homogeneous crystallization requires the formation of a nucleation seed. The nucleation seed creates a surface against the surrounding liquid. This surface makes the existence of the nucleation seed energetically unfavorable for small seed volumes. Therefore, in order for homogeneous crystallization to occur, the system has to overcome a free energy barrier. Consequently, the formation of a nucleus of sufficient size happens with low probability in unbiased MD simulations. This makes the study of crystallization processes difficult and often computationally infeasible. 

In this section, we show that the maximum relative entropy formalism can be used to overcome this free energy barrier and mediate a crystallization of the liquid water system into the hexagonal ice phase I\textsubscript{h}. 
To this end, we simulated an isothermal-isobaric system of water molecules in the hexagonal ice phase I\textsubscript{h} at \SI{10}{K} and \SI{1}{atm}. The TIP4P/ICE model was used as water potential \cite{abascal_potential_2005}. Based on this system, the target RDF $\tilde{g}^{\rm target}_{b}$ was calculated from \SI{0}{\angstrom} to \SI{14}{\angstrom} with a bin size of \SI{0.07}{\angstrom}. By using the computed RDF as input to the maximum relative entropy algorithm, a simulation of liquid water at \SI{360}{K} and \SI{1}{atm} was biased.

A temperature above the freezing point of water at \SI{273}{K} was chosen to ensure sufficient mobility of water molecules in the system. Nevertheless, a phase transition to the hexagonal ice phase was possible in the simulation, since the free energy of the crystalline state is lowered by the induced tilt of the free energy surface (see Sec.~\ref{sec:freeenergy}). 
The Lagrange parameters were updated every 1100 time steps with a step size of \SI{1}{fs}. The step size was $\gamma=2.5$ and no virial weighting ($\kappa=1.0$) was applied. The RDF of the system was computed every 10 time steps and averaged over 1000 time steps to determine the difference between the target RDF and the instantaneous system RDF for the gradient descent update. After the bias update, the system was equilibrated for 100 time steps before next samples of the instantaneous RDF $g_b$ were taken from the simulation. 

After a certain amount of induced bias strength, we observed a gradual increase in temperature in the system during the simulation as the bias was further increased. The rise in temperature originates from the sharp peaks of crystalline RDFs, such as those for water ice at \SI{10}{K}, which result in Lagrange parameters $\lambda_b$ with high absolute values for some bins. In turn, strong forces originate from the large Lagrange parameters $\lambda_b$. As a result, the temperature rise is caused by the high forces that are not adequately damped by the applied thermostat. We were able to mitigate this effect by increasing the damping parameter of the applied thermostat or by decreasing the time steps. However, for moderate temperature increases, we did not observe any problems with reproducing the target RDF. 

The upper part of Figure \ref{fig:ice} shows snapshots of the oxygen atoms contained in the water molecules generated with VMD \cite{humphrey_vmd_1996} during the simulation at time steps \SI{0}{\nano\second} (a), \SI{7}{\nano\second} (b), \SI{8}{\nano\second} (c), \SI{9.5}{\nano\second} (d), and \SI{14}{\nano\second} (e). The oxygen atoms are colored in dependence of a calculated average local bond order parameter $\bar q_6$ \cite{lechner_accurate_2008}, which is a modification of the local bond order order parameter \cite{steinhardt_bond-orientational_1983}, with a cut-off value of \SI{3.5}{\angstrom}. Blue oxygen atoms have an average local bond order parameter $\bar q_6<= 0.07$, while red oxygen atoms are $\bar q_6> 0.07$ (see SI for more information on bond order parameters).

At the beginning of the simulation, the dissimilarity of the RDF of the liquid water and the hexagonal ice system was relatively large, with an MAE of about 0.45. The dissimilarity is also evident from the inlet, which shows the RDF of the unbiased liquid system (blue) and the target RDF of the hexagonal ice (black). 
During the initial time period from \SI{0}{\nano\second} (a) to \SI{7}{\nano\second} (b), the biased simulation exhibited a decrease in density with increasing bias strength. At the same time, the MAE between the target RDF and the RDF originating from the biased system decreased linearly. After approximately \SI{7}{\nano\second} (b), a nucleation seed formed in the biased system. This nucleation seed grew in the following simulation steps. After approximately \SI{10}{\nano\second}, the whole system transitioned into a crystalline phase of I\textsubscript{h} ice (e). The phase transition is visible as a steep drop in MAE between the target and instantaneous RDF.

While some defects were visible in the final solid crystalline structure obtained, the solid phase resembled the structure of hexagonal ice I\textsubscript{h}. The presence of the I\textsubscript{h} ice phase is emphasized by the similarity between the RDF of the biased system after \SI{14}{\nano\second} and the target RDF, as shown in the inset.

\subsection{\label{sec:polymorphs}Polymorphs of \texorpdfstring{\ce{TiO2}}{TiO2}: Rutile and Anatase}

\begin{figure*}[!t]
\centering
    \includegraphics[width=1\linewidth]{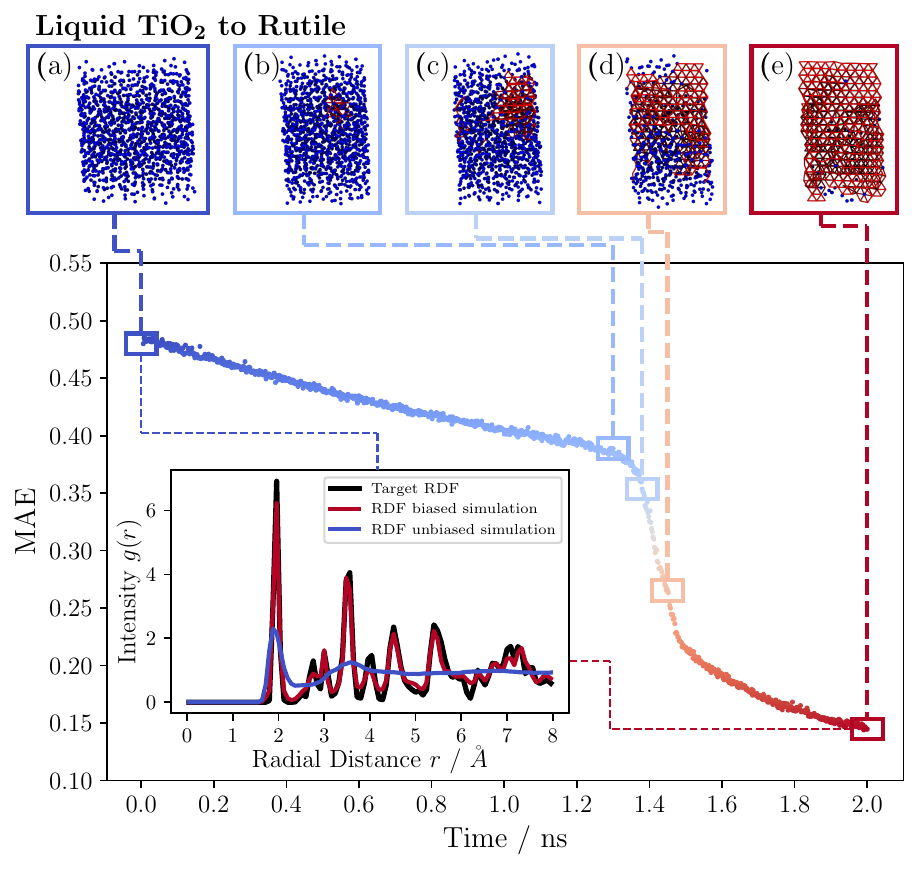}
    \caption{Crystallization of liquid \ce{TiO2} to rutile at \SI{2500}{\kelvin} and \SI{1}{atm}. The mean absolute error (MAE) between the RDF of the biased water system and the target RDF of the simulated rutile is depicted. Several snapshots of the biased simulation, show the state of the titanium atoms in the system from left to right at \SI{0}{\nano\second}, \SI{1.3}{\nano\second}, \SI{1.38}{\nano\second}, \SI{1.45}{\nano\second}, and \SI{2}{\nano\second}. The atoms in the liquid and rutile phases are colored blue and red, respectively. The inlet shows the RDF of the unbiased liquid simulation (blue) and the final state of the system (red) against the target RDF (black).
    }
    \label{fig:rutile}
\end{figure*}

\begin{figure*}[!t]
    \centering
    \includegraphics[width=1\linewidth]{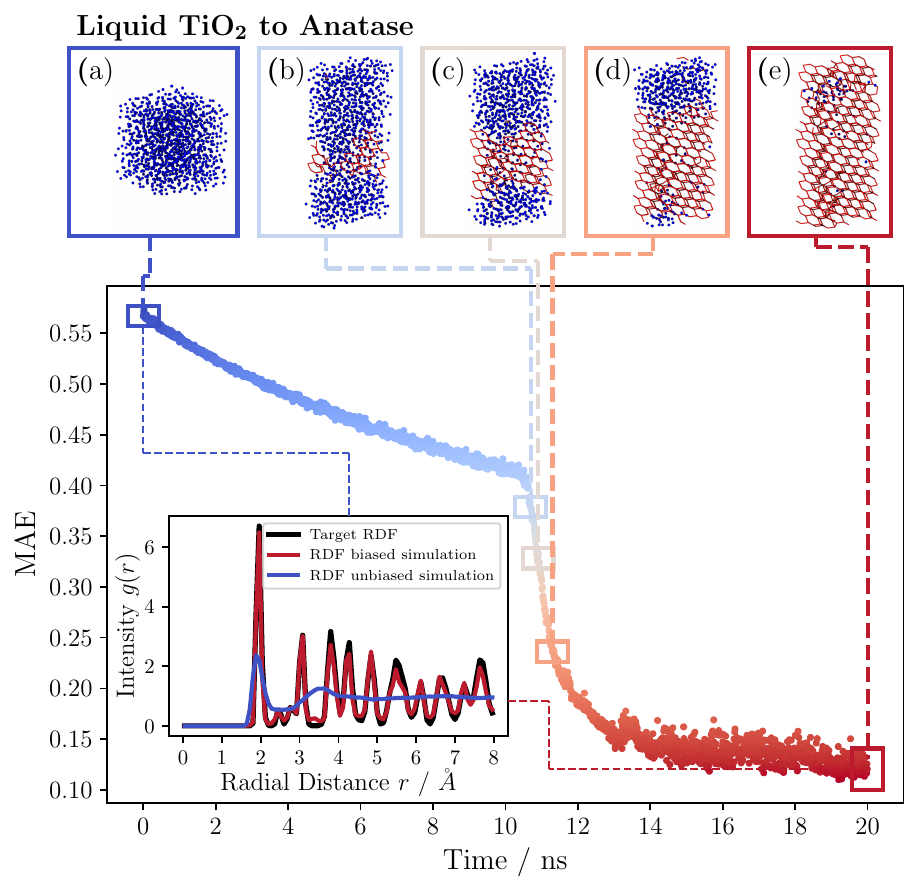}
    \caption{Crystallization of liquid \ce{TiO2} to anatase at \SI{2500}{\kelvin} and \SI{1}{atm}. The mean absolute error (MAE) between the RDF of the biased water system and the target RDF of the simulated anatase is depicted. Several snapshots of the biased simulation, show the state of the titanium atoms in the system from left to right at \SI{0}{\nano\second}, \SI{10.7}{\nano\second}, \SI{10.9}{\nano\second}, \SI{11.3}{\nano\second}, \SI{20}{\nano\second}. The atoms in the liquid and anatase phases are colored blue and red, respectively. The inlet shows the RDF of the unbiased liquid simulation (blue) and the final state of the system (red) against the target RDF (black).
    }
    \label{fig:anatase}
\end{figure*}

\ce{TiO2} is an important material with various applications, such as photocatalytic water splitting \cite{miyoshi_water_2018,hanaor_review_2011}, photodynamic therapy for cancer treatment, inactivation of antibiotic resistant bacteria \cite{ziental_titanium_2020}, and gas sensing \cite{bai_titanium_2014}, to name a few examples.

At atmospheric pressure, there are three different polymorphs of \ce{TiO2}, namely rutile, anatase, and brookite, with rutile and anatase being the most important polymorphs for applications. While rutile is the lowest energy configuration under ambient conditions, anatase and brookite are metastable polymorphs.

In this section, we demonstrate the crystallization of liquid \ce{TiO2} into the corresponding stable rutile or metastable anatase polymorph by biasing a liquid \ce{TiO2} system with the RDF extracted from a rutile or anatase system. 
In particular, the crystallization into the metastable anatase phase would not be feasible using conventional methods developed to find the configuration of the global energy minimum, such as simulated annealing or replica exchange methods.

To bias the liquid \ce{TiO2} system, two target RDFs were created for rutile and anatase, ranging from \SI{0}{\angstrom} to \SI{8}{\angstrom} with a bin size of \SI{0.08}{\angstrom}. 
For this purpose, rutile and anatase were modeled as an isothermal-isobaric ensemble at a temperature of \SI{300}{K} and a pressure of \SI{1}{atm} with the atomic interaction potential developed by Matsui and Akaogi \cite{matsui_molecular_1991} and provided by the Knowledgebase of Interatomic Models (KIM) \cite{tadmor_potential_2011}. The calculated RDFs were used as inputs to the maximum relative entropy algorithm to bias the two identical initial systems of liquid \ce{TiO2} at \SI{2500}{K} and \SI{1}{atm} to reproduce the RDF of either rutile or anatase. A time step size of \SI{1}{fs} was set for both simulations. The Lagrange parameters were updated every 1000 time steps with a step size of $\gamma=5.0$ for rutile and $\gamma=1.25$ for anatase.

Similar to the crystallization of water, we observed an increase in temperature at high bias forces. It is possible to mitigate the temperature increase by increasing the thermostat damping or by using smaller simulation time steps. However, in the present case, the temperature increase did not have a negative effect on the simulations. To counteract the excessive contraction of the system due to the virial contribution $T_{\rm ME}$ of the bias (eq~\eqref{eq:virial}), the update rate of all bins contributing negatively to the total virial contribution was reduced by a factor of $\kappa_b=0.25$. The system RDF was computed at 10 time step intervals and averaged after 1000 time steps to calculate the difference between the target RDF and the instantaneous system RDF for the gradient descent update. It was determined that an equilibration period following the Lagrangian parameter update was unnecessary, as the change in bias potential was sufficiently small in each iteration.

Figure~\ref{fig:rutile} and Figure~\ref{fig:anatase} show the results for the simulations biased with the RDF of rutile and anatase, respectively. In both figures, snapshots of the \ce{Ti} atoms in the biased \ce{TiO2} systems are depicted in the upper part of the figures. The titanium atoms are colored as a function of the averaged local bond order parameter values.
In Figure~\ref{fig:rutile}, blue atoms in the liquid phase correspond to an averaged local bond order parameter $\bar q_8$, calculated over the 12 nearest neighbors, with a value $<= 0.073$, while red atoms in the rutile phase have a value $> 0.073$ (see SI for more information).
Similarly, in Figure~\ref{fig:anatase}, blue atoms correspond to \ce{Ti} atoms in the liquid phase with an averaged local bond order parameter of degree $\bar q_{10}$, calculated over the 12 nearest neighbors, with values $<= 0.085$. Red atoms with $> 0.085$ belong to the anatase phase (see SI for more information).
Both figures show the MAE between the respective target RDF and the instantaneous RDF in the biased system.
Additionally, the RDF of the liquid \ce{TiO2} at the beginning of the biased simulations (blue), the final RDF at the end of the biased simulations after \SI{2}{ns} for rutile and \SI{20}{ns} for anatase (red), and the respective target RDFs (black) are shown in the inlet.

Overall, the figures indicate that both biased simulations proceeded in a similar manner. During the initial phase of the biased simulation, the simulation box stretched in one direction, but the overall density of the system increased.
The initial phase of the biased simulation was characterized by a linear reduction of the MAE in both systems. After the initial phase, a nucleation seed formed in the biased simulations after approximately \SI{1.3}{ns} for the rutile system and \SI{10.7}{ns} for the anatase system. 
From this point on, the system entered a phase transition process and complete crystallization occurred from the nucleation seed. The phase transition was characterized by a steep decrease of the MAE.
After about \SI{1.5}{ns} and \SI{12.0}{ns}, respectively, the transition was complete and no further significant changes in atomic configuration occurred. Except for some visible defects, the systems crystallized completely. The defects were responsible for the small deviations from the final system RDF (red) and the target RDF (black) in the inlet. Overall, the target RDF and the instantaneous system RDF exhibited strong agreement.

The comparative examples of rutile and anatase crystallization demonstrate that the maximum relative entropy formalism can be effectively applied to reproduce different crystal configurations of polymorphs. Furthermore, in this case, the crystallization process was feasible even with minor changes in the simulation or algorithm parameters. More specifically, to reproduce anatase instead of rutile, it was only necessary to change the step size, $\gamma$, and the target RDF.

\section{\label{sec:conclusion}Conclusion and Outlook}
An algorithm was developed to bias MD simulations to reproduce target radial distribution functions. This was done based on the principle of maximum relative entropy. For this purpose, the original atomic interaction potential was linearly biased with a target RDF. The linear functional form of the applied bias, as determined by the maximum entropy approach, is advantageous because it constrains the reproduced ensemble average solely by the information contained in the target RDF.

The magnitude of the bias is determined by the values of the Lagrangian multipliers, which are calculated during the simulation using the gradient descent method. Since the $\lambda_b$ are not uncorrelated observables of the system, it cannot be guaranteed that the optimization function has a global minimum. However, this has not been found to affect the systems studied. The algorithm allows individual atomic contributions to the RDFs to be weighted by atomic scattering factors. This allows the bias of molecular dynamics simulations with RDFs derived from WAXS techniques. The algorithm has been implemented in LAMMPS and evaluated for two use cases.

In one use case, the RDF of a TIP4P/2005 liquid water system was reproduced in a TIP3P liquid water system. The method proved to be effective in accurately reproducing the RDF of the TIP4P/2005 water system while improving the agreement between the angular distribution of the oxygen molecules in the bias and target systems. Based on these results, we can conclude that the approach represents a theoretically sound methodology for updating an interatomic potential using prior knowledge of a systems RDF.

As a second application, the method was used to induce homogeneous crystallization in liquid systems. To achieve this, liquid water was biased to crystallize into hexagonal ice and liquid \ce{TiO2} was biased to crystallize into rutile or anatase. The successful phase transitions demonstrated the ability of our algorithm to reproduce the structure of various crystalline systems and polymorphs with minor imperfections. During the execution of the biased simulations with target RDFs derived from crystalline structures, a temperature increase was observed in strongly biased systems. This increase can be attributed to the sharp peaks of the crystalline RDFs, which in turn lead to high absolute values of the Lagrange parameters $\lambda_b$ for some bins. Consequently, strong forces are added to the simulated system. For moderate temperature increases, no resulting problems were found in the reproduction of the target RDF. Additionally, the effect can be mitigated by increasing the damping parameter of the used thermostat or by downscaling the time steps to shorter lengths.

The method presented here has a variety of potential applications. One such application could be the improvement of liquid interaction potentials by biasing the potential with a RDF, as demonstrated for the TIP3P water model. Another application could be the study of liquid-to-solid phase transitions. Through a sophisticated combination of biased and unbiased simulations, it may be possible to prepare states along a crystallization transition path and continue with unbiased simulations from these states, or use the prepared states as starting points for advanced sampling methods, such as umbrella sampling.

In addition, the algorithm described here could assist in the interpretation of experimentally obtained RDFs from WAXS measurements of liquid or solid states. While the atomic states considered in this work are well known, this algorithm could reveal other atomic configurations by combining measured RDFs with MD simulations. This is particularly true for metastable states, as the method presented here is able to reproduce such metastable states, in contrast to other methods such as simulated annealing. General purpose machine-learned atomic interaction potentials, which can extrapolate energies of stable atomic configurations to which they have not been directly trained \cite{chen_universal_2022,batatia_foundation_2024}, in combination with the algorithm presented here, offer the possibility to interpret WAXS measurements of atomic structures that are still undiscovered.

In the future, the method may also be applied to more complex systems, such as crystallization processes of confined matter, where only the confined atoms would be distorted by the measured RDFs.

The method can also be used to provide atomic configurations for training machine learning potentials. For a machine learning potential to interpolate effectively, a diverse range of atomic configurations must be provided as undersampled regions of the configuration space can lead to unsatisfactory generalization of the models. However, generating a sufficient number of atomic configuration samples, especially in the transition regions of a crystallization process, is a challenging task. In this way, our method could contribute to the training of machine learning potential specialized for the study of specific phase transitions.

In addition to the discussed applications, the presented method offers many directions for further research. With respect to the interpretation of WAXS data, it might be interesting to develop a method that uses the principle of maximum relative entropy to bias MD simulations with the directly measurable structure factor function $S(q)$. A bias derived directly from the structure factor function $S(q)$ would avoid an initial Fourier transform of the measurement data, which in turn would avoid additional noise. In general, the application of the principle of maximum relative entropy can be used in conjunction with any system observable derived from the ensemble average. Therefore, the use of local bond order parameters to bias MD systems would be of interest, as these parameters are designed to discriminate between different atomic configurations.

\begin{acknowledgement}

L.D. was supported by the Data Science in Hamburg HELMHOLTZ Graduate School DASHH, Helmholtz Association Grant-No. HIDSS-0002. 
We also acknowledge the scientific exchange and support of the Centre for Molecular Water Science CMWS and the research initiative BlueMat: Water-Driven Materials, Hamburg (Germany).
This research was supported in part through the Maxwell computational resources operated at Deutsches Elektronen-Synchrotron DESY, Hamburg, Germany.
This work was funded by the Deutsche Forschungsgemeinschaft (DFG, German Research Foundation) -- 192346071; 390794421.

\end{acknowledgement}

\begin{suppinfo}

The following file is available free of charge:
\begin{itemize}
  \item SIAverageBondOrderParameters.pdf: Explanatory content regarding the used average local bond order parameters
\end{itemize}
The simulation code is available under:\\
\url{https://collaborating.tuhh.de/m-29/software/maxentrdf}

\end{suppinfo}

\bibliography{references}

\end{document}